\documentclass[aps, prl, reprint, floatfix, superscriptaddress, longbibliography]{revtex4-1}

\usepackage{graphicx}
\usepackage{amsmath}
\usepackage{epstopdf} 
\usepackage[colorlinks=false, pdfborder={0 0 0}]{hyperref} 

\begin{document}

\title{Shot Noise of a Quantum Dot Measured with Gigahertz Impedance Matching}

\author{T. Hasler}
\affiliation{Department of Physics, University of Basel, Klingelbergstrasse 82, CH-4056 Basel, Switzerland}

\author{M. Jung}
\affiliation{Department of Physics, University of Basel, Klingelbergstrasse 82, CH-4056 Basel, Switzerland}

\author{V. Ranjan}
\affiliation{Department of Physics, University of Basel, Klingelbergstrasse 82, CH-4056 Basel, Switzerland}

\author{G. Puebla-Hellmann}
\affiliation{Department of Physics, University of Basel, Klingelbergstrasse 82, CH-4056 Basel, Switzerland}
\affiliation{Department of Physics, ETH Z\"urich, Otto-Stern-Weg 1, CH-8093 Z\"urich, Switzerland}

\author{A. Wallraff}
\affiliation{Department of Physics, ETH Z\"urich, Otto-Stern-Weg 1, CH-8093 Z\"urich, Switzerland}

\author{C. Sch\"onenberger}
\affiliation{Department of Physics, University of Basel, Klingelbergstrasse 82, CH-4056 Basel, Switzerland}

\date{\today}

\begin{abstract}
The demand for a fast high-frequency read-out of high-impedance devices, such as quantum dots, necessitates impedance matching. Here we use a resonant impedance-matching circuit (a stub tuner) realized by on-chip super\-conducting transmission lines to measure the electronic shot noise of a carbon-nano\-tube quantum dot at a frequency close to $3$\,GHz  in an efficient way. As compared to wideband detection without impedance matching, the signal-to-noise ratio can be enhanced by as much as a factor of 800 for a device with an impedance of $100$\,k$\Omega$. The advantage of the stub resonator concept is the ease with which the response of the circuit can be predicted, designed and fabricated. We further demonstrate that all relevant matching circuit parameters can reliably be deduced from power-reflectance measurements and then used to predict the power-transmission function from the device through the circuit. The shot noise of the carbon-nano\-tube quantum dot in the Coulomb blockade regime shows an oscillating suppression below the Schottky value of $2eI$, as well as an enhancement in specific regions.
\end{abstract}

\maketitle

\section{Introduction}
Noise studies are shown to be a powerful tool to characterize electron transport in quantum systems \cite{blanter2000}. They reveal information which is not accessible via the conductance alone. In particular, correlations due to e.g.~quantum statistics or Coulomb repulsion lead to a suppression or enhancement of the nonequilibrium shot noise relative to the classical value given by Schottky, $S_I=2e \lvert I\rvert$. Here, $S_I$ is the current-noise spectral density and $I$ denotes the time-averaged dc current. Correlations can be observed notably in low-dimensional nano\-scale devices due to coherent charge transport and reduced screening by the environment. Quantum dots (QDs), representing one of the smallest systems possible, are currently of particular interest for instance as building blocks for spintronics-based quantum computation \cite{levy2002}.

The trend in modern experiments is toward a fast read-out of QD states using high frequencies. However, the combination of high-frequency measurements with QD impedances on the order of $R=100$\,k$\Omega$ or larger suffers from the large impedance mismatch to the standard line and instrument impedance of $Z_0=50 \Omega$, leading to a strong suppression of the detected signal power on the order of $(Z_0/R)^2$. In order to measure noise of a QD, the noise signal should be efficiently transmitted into the $50$ $\Omega$ line that connects to the amplifier. This can be achieved with an impedance-matching circuit\cite{xue2007, puebla2012, altimiras2014, ranjan2015}. Here, we use a stub impedance-matching circuit consisting of two low-loss super\-conducting transmission lines connected in parallel, with a resonance frequency close to $3$\,GHz. For the presented QD sample, the signal-to-noise ratio (SNR) enhancement at a resistance of $100$\,k$\Omega$ is deduced to be up to a factor of $200$ as compared to a wide-band detection without impedance matching. The upper bound for the improvement in SNR at this resistance is as large as a factor of $800$, assuming a lossless impedance-matching circuit at full matching. The device and matching circuit are placed on the same chip to minimize parasitic capacitances and inductances.

In this work, we use a carbon-nanotube (CNT) QD as a model system to demonstrate the application of the stub impedance-matching circuit for sensitive gigahertz-frequency noise measurements of high-resistance samples. QDs defined in  CNTs are shown to produce well-resolved and stable results \cite{steele2009, waissman2013, jung2013}. Several studies investigate shot noise of CNT QDs in various regimes. There are measurements reported in the co\-tunneling regime \cite{onac2006}, in the Kondo regime \cite{delattre2009, basset2012}, in the transparent case showing Fabry-Perot interferences \cite{wu2007, herrmann2007, kim2007} and at very high bias \cite{wu2010}, where electron-phonon coupling becomes evident. These measurements are performed with broadband detection methods.

\section{Sample fabrication}

\begin{figure}
\includegraphics{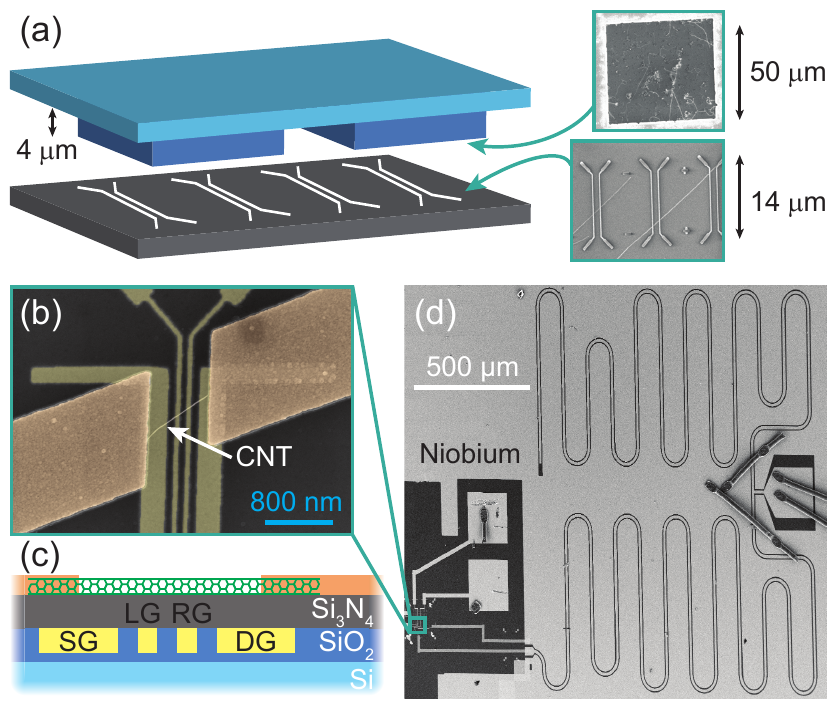}
\caption{\label{fig1} (a) Illustration of the carbon nanotube stamping process. CVD growth is done on the top silicon substrate (blue), which contains an area of square mesas. On the bottom substrate (gray), bottom gates are fabricated and covered with silicon nitride. (b) False-color image of the CNT connected to Ti/Au leads (orange) and bottom gates underneath (yellow), which are covered with silicon nitride. (c) Sketch of the cross section. The gates called source gate (SG), drain gate (DG), left gate (LG) and right gate (RG) are covered with silicon nitride. The CNT is stamped on top and contacted with Ti/Au leads.  (d) SEM image of the stub impedance-matching circuit made out of two niobium coplanar transmission lines in parallel. The 50\,$\Omega$ side is at the launcher on the right, and the high-resistance sample is located at the bottom left. The two bond wires are air bridges to connect the ground planes.}
\end{figure}

The challenge on the fabrication side is to combine a low-disorder CNT with a high-quality microwave circuit. CNT growth is done by chemical vapor deposition (CVD) in a $\mathrm{CH}_4$ and $\mathrm{H}_2$ atmosphere at $950\,^{\circ}\mathrm{C}$ \cite{babic2004}. This process turns out to be harmful to silicon nitride and oxide substrates. Resonators fabricated on these substrates after CVD growth exhibit quality factors below 100 at 4.2 K. That is why we apply a CNT stamping technique adapted from Viennot, Palomo, and Kontos \cite{viennot2014}, which is sketched in Fig.~\ref{fig1}(a). In contrast to single-CNT stamping with a fork \cite{muoth2013, ranjan2015}, we transfer many CNTs from a growth substrate to an area on the target substrate where bottom gates have been fabricated. Selected CNTs are then contacted.

In more detail, the fabrication steps are as follows: We use a silicon substrate with a thermal oxide for the stamps. It is patterned within a $2\times2 \, \mathrm{mm}^2$ area into an array of square mesas. The squares with a lateral size of 50 $\mu$m and a height of 4 $\mu$m [top right of Fig.~\ref{fig1}(a)] are separated by 50 $\mu$m. After spinning Fe/Mo catalyst particles onto the SiO$_2$, we do CVD growth. On a target Si/SiO$_2$ substrate, an array of Au bottom gates is deposited and covered with a 50-nm layer of plasma-enhanced CVD silicon nitride. With the help of a mask aligner, the stamp and the bottom-gate area of the target substrate are roughly aligned on top of each other and then pressed together. Then we locate CNTs that are crossing bottom gates, as in the bottom right of Fig.~\ref{fig1}(a), using a scanning electron microscope (SEM). Next, the silicon nitride layer covering the gates is etched with a $\mathrm{CHF_3/O_2}$ plasma some distance away from the corresponding CNT. In the following, we contact the CNT and its gates with 5/40 nm-thick Ti/Au leads. The resulting device is shown in Fig.~\ref{fig1}(b), and a cross section is sketched in Fig.~\ref{fig1}(c). The next fabrication step is to protect the CNT and the gates with a PMMA/HSQ bilayer resist before a 150-nm-thick layer of niobium is sputtered. Subsequently, the niobium on the protected area is lifted off, and the impedance-matching circuit, as shown in Fig.~\ref{fig1}(d), is patterned by electron beam lithography using a PMMA resist layer and reactive-ion etching with an $\mathrm{Ar/Cl_2}$ plasma. The final device is glued into the center of a printed circuit board with rf and dc connectors, which is our sample holder. The ground planes on the sample holder and on the chip are connected with many bond wires located at the edge of the chip (not visible). In order to suppress any spurious electro\-magnetic modes arising from different potentials on the ground planes, Al bond wires connect the ground planes close to the T junction near the $50$\,$\Omega$ launcher, seen in Fig.~\ref{fig1}(d) on the right side.

\section{High-frequency setup}
\begin{figure}
\includegraphics{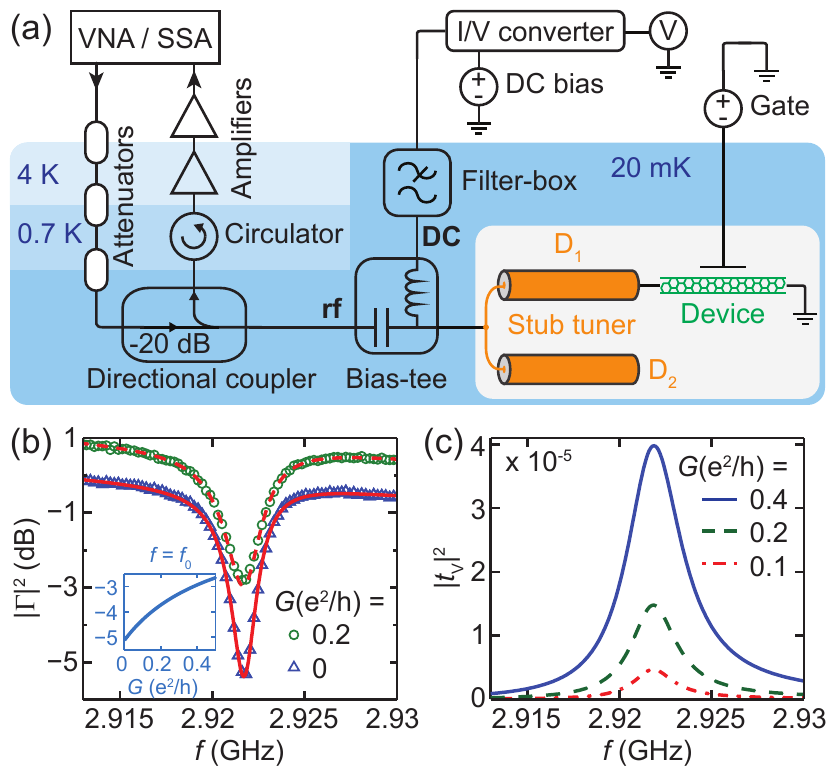}
\caption{\label{fig2} (a) Schematic of the setup with an input and an output rf line plus one dc line. Everything on a blue background is inside a dilution refrigerator with a base temperature of 20\,mK. (b) Amplitude squared of the reflection coefficient $\Gamma=V_{\mathrm{out}}/V_{\mathrm{in}}$ around the resonance frequency. Symbols are measured and lines fitted or calculated. The stub tuner loss $\alpha=0.046\,\mathrm{m^{-1}}$ as well as the two lengths $D_1=10.355$\,mm and $D_2=10.589$\,mm are extracted by fitting (solid red line) to the spectrum in the Coulomb blockade regime of the QD, where $G=0$ (blue triangles). The upper spectrum for a finite dc conductance of $G=0.2\,e^2/h$ is plotted with a shift of 1 dB for clarity (green circles). It matches well the calculated reflection coefficient (dashed red line) using the previous fit parameters. The conductance dependence of the reflectance at the resonance frequency is plotted in the inset. (c) Calculated voltage-transmission function of the stub tuner for three typical device conductances using the fit parameters gained from (b).}
\end{figure}

The resulting sample is measured at 20 mK by using the setup depicted in Fig.~\ref{fig2}(a). An attenuated rf line can be used to apply a signal to the sample for reflectometry. The applied power at the sample is about $-122$\,dBm. The reflected signal - or emitted noise - is separated from the input line via a directional coupler. A circulator prevents room-temperature radiation from reaching the sample through the output line. The outgoing signal is fed into a cryogenic and a room-temperature amplifier and detected with a vector network analyzer (VNA) for reflectrometry or a signal and spectrum analyzer (SSA) for noise measurements. Via a bias tee, one can add a dc bias and record the dc current simultaneously with rf measurements. Thermal radiation propagating through the dc cables is filtered with homemade silver-epoxy microwave filters \cite{scheller2014}.

The key part of the experiment is the impedance-matching circuit called the stub tuner, which has previously shown to enable sensitive reflectometry \cite{hellmueller2012, puebla2012, ranjan2015}. It consists of two coplanar transmission lines (CTLs) in parallel, with lengths $D_1$ and $D_2$ close to $\lambda_0 /4$ ($\lambda_0$ being the wavelength at resonance). One CTL end ($D_1$) is connected to the device with conductance $G$, whereas the other end ($D_2$) is open-ended. The commercial software Sonnet is used to conduct electromagnetic simulations of the CTL. The extracted geometrically defined characteristic impedance is $Z_0^*=44.8 \, \Omega$ and the effective dielectric constant $\epsilon_\mathrm{eff}=6$. The input admittance $Y_\mathrm{in}$ of the stub tuner is the sum of the two arm admittances and reads \cite{ranjan2015, pozar2005}
\begin{equation}
Y_\mathrm{in} = \frac{1}{Z_0^*} \left(\frac{Z_0^* G+\tanh\left(\gamma D_1 \right)}{1+Z_0^* G \tanh\left(\gamma D_1 \right)} + \tanh\left(\gamma D_2 \right) \right),
\end{equation}
with the propagation constant $\gamma=\alpha+\mathrm {i} \cdot \beta$ containing the loss $\alpha$ and the wave number $\beta=2 \pi f \sqrt{\epsilon_\mathrm{eff}}/c$, where $f$ is the frequency, $\epsilon_\mathrm{eff}$ the effective dielectric constant and $c$ the speed of light. The reflection coefficient (ratio of reflected to incoming voltage) when connected to a line with characteristic impedance $Z_0$ is then given by
\begin{equation}
\label{eq_Gamma}
\Gamma = \frac{e^{\mathrm{i} \phi}-Z_0 Y_\mathrm{in}}{e^{\mathrm{i} \phi}+Z_0 Y_\mathrm{in}},
\end{equation}
where the phase factor $e^{i\phi}$ is a fit parameter that accounts for the asymmetry in the resonance caused by standing waves in the setup.

The strategy is now as follows: One first deduces the stub tuner parameters from a frequency-dependent power-reflectance measurement for a known conductance value $G$ of the CNT device. Once all the parameters in the matching circuit are fixed, one can use them to determine $G$ for an arbitrary gate setting. We use a gate setting deep in the Coulomb blockade (CB) regime, where $G=0$, as a reference to deduce the stub tuner parameters $D_1$, $D_2$, and $\alpha$ by fitting the measured $\lvert \Gamma (f) \rvert^2$ with Eq.~(\ref{eq_Gamma}) as shown in Fig.~\ref{fig2}(b). To demonstrate that this extraction works reliably, the measured spectrum at a dc conductance $G=0.2 \, e^2/h$ is compared in the figure to the spectrum calculated with the previously determined fitting parameters. An excellent agreement is evident. This demonstrates that one can now use this procedure to determine the differential high-frequency $G$ for any gate-voltage setting by reflectometry, i.e. by fitting to the measured reflected power. The inset in Fig.~\ref{fig2}(b) shows the dependence of the reflectance on $G$ at the resonance frequency. Full matching is not reached in this sample. The reflection dip is deepest for $G=0$ and decreases with increasing conductance. Hence, it is even possible to infer the conductance $G$ just by measuring the resonance amplitude.

For noise measurements, we need to know the voltage transmission through the stub tuner from the sample to the $50$\,$\Omega$ side. It can be calculated with the stub parameters obtained from reflection. Solving the wave equation of a stub tuner with the appropriate boundary conditions at the two ends and at the T junction between the two arms and the launcher (derivation in Supplemental Material \cite{supp}), one obtains a voltage-transmission function
\begin{equation}
\label{eq_tv}
t_V = \frac{Z_0}{R+Z_0} \cdot \frac{2 e^{\gamma D_1} \coth \left(\gamma D_2 \right)}{\Gamma_L + e^{2 \gamma D_1} \left[ 1 + 2 \coth \left( \gamma D_2 \right) \right]},
\end{equation}
with the differential device resistance $R=1/G$, $\Gamma_L = \left( R-Z_0 \right) /  \left( R+Z_0 \right)$ and assuming that $Z_0^* \approx Z_0$. The resulting power transmission with the previously determined stub parameters is plotted in Fig.~\ref{fig2}(c). The stub tuner has a bandpass effect around the resonance frequency $f_0$. The bandwidth, defined as full width at half maximum, can be inferred to be $\mathrm{BW_{stub}} = f_0 \cdot 4 Z_0 G / \pi$ at matching in the limit where the loss $\alpha \ll 1$. In our case, we obtain a bandwidth of 1.5 MHz for $G=0.2 \, e^2/h$ (corresponding to $R = 130$\,k$\Omega$).

We measure the amplified noise power over $Z_0=50 \, \Omega$ integrated over a bandwidth (BW) of 20 MHz around the resonance frequency, defined as $\langle \delta P \rangle$. For each gate voltage, the corresponding background noise $\langle \delta P_0 \rangle$ at $V_\mathrm{SD}=0$, containing amplifier and thermal noise, is subtracted. By dividing the setup amplification $g$ as well as the stub tuner transmission $t_V$ [Eq.~(\ref{eq_tv})] integrated over the bandwidth and converting power to current, the resulting shot-noise spectral density is (see Supplemental Material for details \cite{supp})
\begin{equation}
\label{eq_si}
S_I = G^2 Z_0 \frac{\langle \delta P \rangle - \langle \delta P_0 \rangle}{g \int_\mathrm{BW} \lvert t_V\rvert ^2 \, df }.
\end{equation}

The setup power gain $g$ appearing in Eq.~(\ref{eq_si}), which includes the gain of the amplifiers and cable loss, is determined by replacing the sample with a metal-wire resistor in the hot-electron regime. In this regime, the Fano factor $F=S_I/2eI$ is $\sqrt{3}/4$ \cite{steinbach1996, henny1998}. The wire length of $L=50 \, \mu$m is longer than the inelastic electron scattering length but shorter than the electron-phonon scattering length. Its width of 680 nm and thickness of 30 nm lead to a residual resistance of 39 $\Omega$, which is close enough to 50\,$\Omega$ to have a high signal output without impedance matching. The wire is attached to two copper pads of size 300 $\times$ 300 $\mu \mathrm{m}^2$ and height 500 nm, acting as heat sinks. Comparing the shot-noise dependence on current in the linear regime with the Fano factor $\sqrt{3}/4$, we can infer a power gain $g=97.9$\,dB of the amplification chain (see Supplemental Material \cite{supp}).

\section{Experiment}

\begin{figure}
\includegraphics{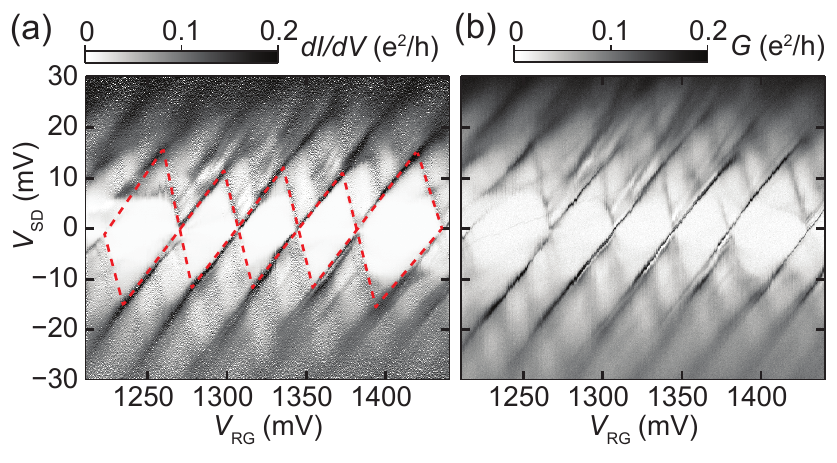}
\caption{\label{fig3} (a) Derivative of the dc current ($dI/dV_\mathrm{SD}$) as a function of the voltage on the right gate and of the source-drain bias. The contour of the CB diamonds is highlighted by the dashed line. (b) Differential conductance deduced from the reflection amplitude.}
\end{figure}

With the help of the bottom gates beneath the CNT, we conduct measurements in the single-QD regime (more information on the QD formation can be found in Supplemental Material \cite{supp}). Its energy levels are controlled by the plunger gate voltage $V_\mathrm{RG}$ \cite{nygard1999}. The differential conductance derived by numerically differentiating the measured dc current ($dI/dV_\mathrm{SD}$) and by transforming the reflection amplitude to $G$ using Eq.~(\ref{eq_Gamma}) is shown in Figs.~\ref{fig3}(a) and \ref{fig3}(b), respectively. The comparison shows that the rf conductance is in good agreement with $dI/dV_\mathrm{SD}$, which confirms the validity of our way to extract the stub tuner parameters. The fourfold degeneracy of the CNT QD states becomes evident by looking at the dashed contour lines. We stress that the rf-deduced conductance is in fact less noisy and can be measured much faster.

\begin{figure}
\includegraphics{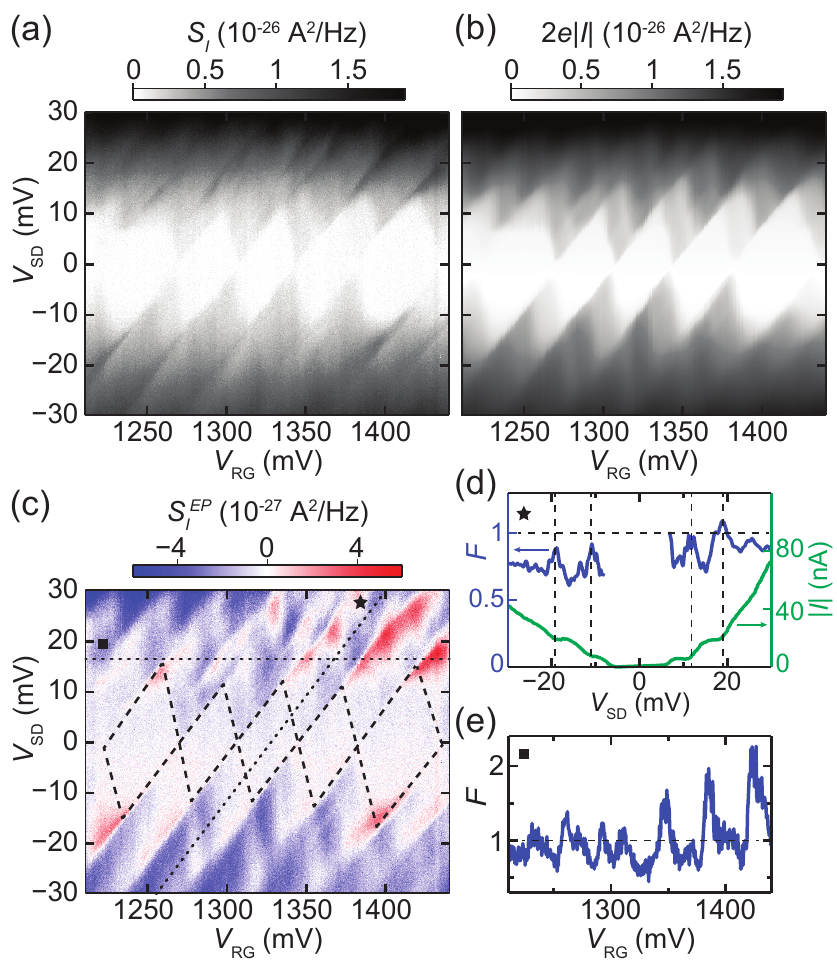}
\caption{\label{fig4} (a) Calibrated shot-noise spectral density $S_I$ as a function of voltage on the right gate and of source-drain voltage. (b) Schottky noise $2e\lvert I\rvert$ and (c) excess Poissonian noise $S_I^\mathrm{EP}=S_I-2e\lvert I\rvert$. The CB diamond contours (dashed lines) are copied from the conductance plot in Fig.~\ref{fig3}(a). (d) Fano factors averaged over a range of 1.2 mV in $V_\mathrm{SD}$ (left scale) and absolute value of current (right scale) along the dotted lines in (c) marked with a star. Fano factor peaks correspond to the onset of current transitions from one plateau to the next. (e) Fano factors along the horizontal line in (c) (marked with a square) exceeding one.}
\end{figure}

Figure \ref{fig4}(a) shows the current noise $S_I$ measured in the same gate range after applying Eq.~(\ref{eq_si}) for calibration. To compare with the Schottky noise, we show in Fig.~\ref{fig4}(b) a calculated plot of $2 e \lvert I \rvert$, where $I$ represents the measured dc current. The so-called excess noise, which is the difference $S_I^\mathrm{EP} = S_I - 2 e \lvert I \rvert$, is shown in Fig.~\ref{fig4}(c). One can distinguish between super-Poissonian noise, where $S_I^\mathrm{EP}$ is positive and sub-Poissonian noise, where it is negative. Inside the CB regime, namely at the corners of the two large diamonds, there are some small areas where the noise is super-Poissonian. This shot-noise enhancement might originate from bunched charge transport due to inelastic cotunneling \cite{sukhorukov2001, belzig2004, belzig2005, onac2006, zhang2007, gustavsson2008, okazaki2013}.

Outside the CB, an oscillating shot-noise reduction is apparent. More insight is gained in Fig.~\ref{fig4}(d), where the Fano factor along the diagonal dotted line in Fig.~\ref{fig4}(c) is plotted together with the absolute value of the current. The current plateaus in the Coulomb staircase coincide with a peak in the Fano factor with $F \sim 1$, whereas the shot noise is suppressed at the transitions from one plateau to the next. This behavior can be explained by a simplified model that takes only next-nearest charge states into account, i.e.~\ charge states $N$ and $N+1$ \cite{hershfield1993, birk1995}. The observation of a Coulomb staircase is a sign that the bare tunneling rates of the two junctions are quite different. As a consequence, the current is mostly determined by the more opaque tunnel junction yielding a Fano factor close to one. This is the case on the current plateaus where the charge state is fixed to one charge value for most of the time. In contrast, at the transition between two current plateaus, the two corresponding charge states $N$ and $N+1$ are equally probable. This is caused by a subtle energy dependence of the effective tunneling rates that takes the charging energy of the island into account \cite{hershfield1993}. Hence, in this case the whole device behaves as if it was composed of two identical junctions in series with similar tunneling rates. This yields a suppression of the Fano factor by 2 to $F=0.5$ in the ideal case. However, at finite temperature and/or for larger bias voltages, more than two charge states are involved, yielding $F > 0.5$. The periodic noise suppression therefore tends to decay away at large bias voltages and approaches $F=1$ for $eV \gg E_c$. This is exactly what we see in the data.

But it can be seen in Fig.~\ref{fig4}(d) and more pronouncedly in Fig.~\ref{fig4}(e) that the Fano factor peak values can exceed one. This indicates that the assumption of the two-state model used above is too much simplified and there is more than one channel involved. If, for example, two different orbital states are accessible within the bias window, it is very likely that their lead couplings are different. Sequential tunneling may then be rapid through the strongly coupled orbital until this process is interrupted when an electron is trapped in the weakly coupled state. This results in a random sequence of electron bunches with a noise that exceeds the classical Schottky value \cite{belzig2005, gustavsson2006}.

\section{Conclusion and discussion}

In summary, we demonstrate the versatility of a matching circuit realized by a stub tuner for quantitative noise measurements of high-impedance quantum devices at gigahertz frequencies. Our model system for a quantum device is a single CNT QD. The CNT is transferred from a growth chip to the device chip by stamping. The simple planar structure of a stub tuner built from coplanar transmission lines makes it easy to design and to fabricate with standard lithography. We show that all relevant circuit parameters can be deduced from the reflection spectrum. These parameters can then be used to calculate the transmission function needed to quantify the noise spectral density of the device.

In order to quantify the advantage in noise measurement due to the matching circuit, one has to compare with a wideband noise detection without any impedance matching. The later case offers significantly more BW, but the noise signal is strongly reduced due to impedance mismatch. But, in practice, the BW is not infinite but limited by the circulator and amplifier to values in the range of about 500 MHz. The total power $\langle \delta P_2 \rangle$ in the detection line before amplification is given by $1/2 \cdot S_I Z_0 f_0$ and $S_I Z_0 \mathrm{BW}$ for the case with and without matching circuit, respectively (see Supplemental Material \cite{supp}). Here, $f_0$ denotes the resonance frequency of the matching circuit. It is evident that the matching circuit provides an improvement in detected power, since $f_0 > \mathrm{BW}$. 

However, the real strength of the stub tuner is evident only if one also considers the background noise. Narrowband detection greatly reduces the collected background noise added by the setup, for example, by the amplifier. This is captured by the SNR, which is defined as the desired noise signal divided by the background noise, where the background noise is due to the amplifier chain. Here, the main results of the SNR analysis are given; the derivations can be found in Supplemental Material \cite{supp}. In order to compare the efficiency of a matching circuit for different matching conditions and even different impedance-matching circuits, we introduce the figure of merit $g_\mathrm{SNR}=\mathrm{SNR_{matching}}/\mathrm{SNR_{no\,matching}}$, which is given by
\begin{equation}
g_\mathrm{SNR}=\left( \frac{R}{Z_0} \right) ^2 \cdot \frac{\int_\mathrm{BW} \lvert t_V\rvert ^2 \, df}{\mathrm{BW}}.
\end{equation}
The figure of merit depends on the device resistance $R$, the circuit bandwidth BW and the transmission function $t_V$. The upper bound in the lossless case at full matching is derived to be 
\begin{equation}
g_\mathrm{SNR}^\mathrm{max}=\frac{\pi}{8}\frac{R}{Z_0} \approx 800 \,\,\, \mathrm{for} \,\,\, R=100\, \mathrm{k\Omega}.
\end{equation}
Despite being quite far from full matching and having some loss in the circuit, the figure of merit for the device presented here is still as high as $g_\mathrm{SNR} \approx 200$ at a device resistance of $R=100$\,k$\Omega$. It is interesting to note that the figure of merit for an $LC$-matching circuit \cite{xue2007, zhang2007, okazaki2013} is exactly the same, although the bandwidth is larger. The bandwidth scales with $\sqrt{Z_0/R}$ compared to $Z_0/R$ for a stub tuner. 

In conclusion, a matching circuit can provide a tremendous increase in performance for noise measurements and other experiments in which a high signal transmission is crucial. The increase in SNR is the same for a stub tuner and an $LC$ circuit. However, the stub tuner circuit can be designed in a much easier manner, but it is of narrower bandwidth. This may be an advantage or disadvantage depending on the application. If spurious resonances appear in the same frequency window, for example, so-called box modes generated by the sample enclosure box, it might be beneficial not to have a too-large bandwidth. If, on the other hand, fast read-out is the key, an $LC$ circuit could perform better. 

\section*{Acknowledgements}

The authors thank T. Kontos, J. J. Viennot and L. Bruhat for their kind help with CNT stamping. For the silicon nitride deposition, we thank D. Marty, C. D. Wild, and J. Gobrecht from the PSI. We acknowledge financial support by the ERC project QUEST and the Swiss National Science Foundation (SNF) through various grants, including NCCR-QSIT.

\end{document}


\title{Supplemental Material for ''Shot Noise of a Quantum Dot Measured with Gigahertz Impedance Matching''}

\author{T. Hasler}
\affiliation{Department of Physics, University of Basel, Klingelbergstrasse 82, CH-4056 Basel, Switzerland}

\author{M. Jung}
\affiliation{Department of Physics, University of Basel, Klingelbergstrasse 82, CH-4056 Basel, Switzerland}

\author{V. Ranjan}
\affiliation{Department of Physics, University of Basel, Klingelbergstrasse 82, CH-4056 Basel, Switzerland}

\author{G. Puebla-Hellmann}
\affiliation{Department of Physics, University of Basel, Klingelbergstrasse 82, CH-4056 Basel, Switzerland}
\affiliation{Department of Physics, ETH Z\"urich, Otto-Stern-Weg 1, CH-8093 Z\"urich, Switzerland}

\author{A. Wallraff}
\affiliation{Department of Physics, ETH Z\"urich, Otto-Stern-Weg 1, CH-8093 Z\"urich, Switzerland}

\author{C. Sch\"onenberger}
\affiliation{Department of Physics, University of Basel, Klingelbergstrasse 82, CH-4056 Basel, Switzerland}


\maketitle
\thispagestyle{plain} 

\onecolumngrid

\begin{figure}[H]
\centering
\includegraphics[width=0.9\textwidth]{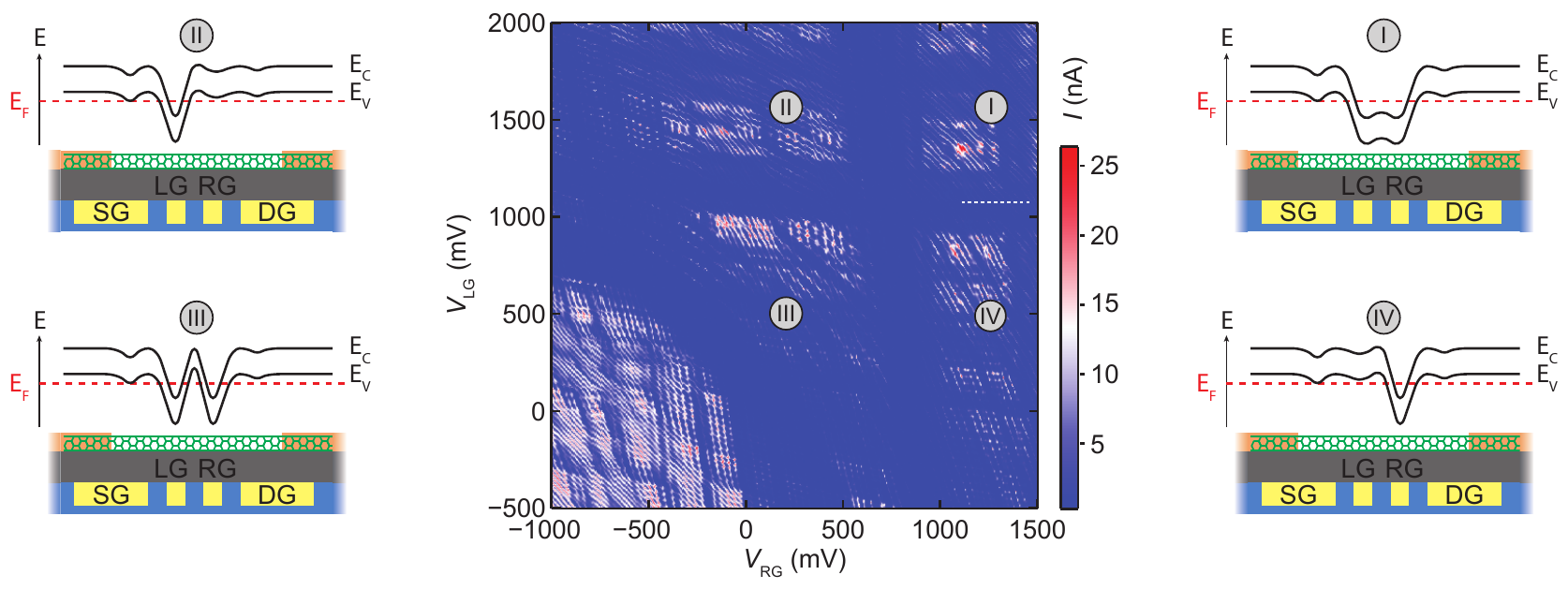}
\captionsetup{width=0.9\textwidth}
\caption{\label{figS1} Main figure: dc current as a function of the voltages on the left gate (LG) and right gate (RG) for a source-drain bias of 1\,mV.}
\end{figure}

\textbf{Quantum dot formation.} 
The dc current on a large range of left gate (LG) and right gate (RG) voltages is plotted in the main figure of Fig.~\ref{figS1}. The source gate (SG) as well as the drain gate (DG) are fixed at $-3$\,V to get p-doped leads. The valence and conduction band edges with respect to the Fermi level are depicted for four different regimes I-IV, which are separated by a band-gap transition. With the help of the two central gates (LG, RG), one can get a single quantum dot (QD) in the centre (I), a QD above the LG (II), a QD above the RG (IV) or a triple QD (III). It is best seen in the left bottom corner, where the entire CNT should be p-doped, that Fermi-level pinning close to the leads can still lead to a QD formation. For the measurements presented in the main text, we concentrate on the single QD regime (I). If plotted with a different color bar, weaker resonance lines would still appear in the figure at the white dashed line, where the left gate is set to $V_\mathrm{LG}=1076$\,mV. Since the couplings to source and drain are weak enough to get a good confinement for clearly visible QDs in this regime, this is chosen to be the working regime for all measurements in the main text. 

The excited state spacing in this single QD regime can be deduced from Fig.~3 in the main text and results in about 8\,meV. It can be compared with the single-particle level spacing $\Delta E = 1 \, \mathrm{meV} / L \, \mathrm{(\mu m)}$ \cite{nygard1999}. The obtained QD length is $L=130$\,nm, meaning that the QD is located mainly between the two central gates, which are separated by 100\,nm.

\begin{figure}[H]
\centering
\includegraphics{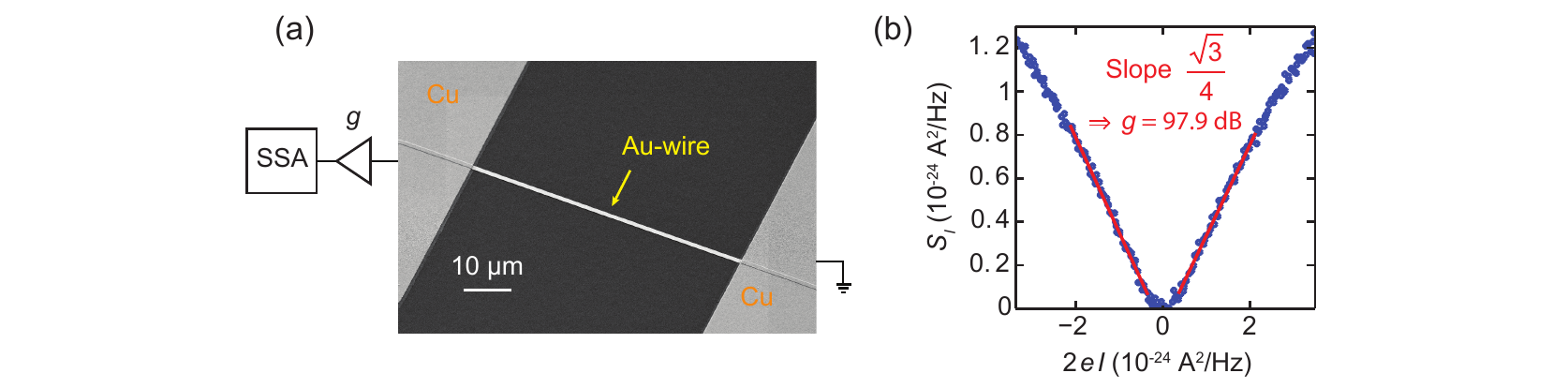}
\captionsetup{width=0.9\textwidth}
\caption{\label{figS2} (a) Measurement setup for the amplifier calibration with an SEM image of the gold wire between the two copper pads. (b) Shot noise as a function of the full Schottky noise $2 e I$ with subtracted background noise, consisting of thermal and amplifier noise.}
\end{figure}

\textbf{Setup calibration.} 
For the calibration of the setup, the sample is replaced by a metal wire with a well-known noise emission [see Fig.~\ref{figS2}(a)]. The gold wire is 30 nm-thick and 680 nm-wide with a residual resistance of 39\,$\Omega$. It is connected to two 500 nm-thick copper pads of the size $300 \times 300 \, \mathrm{\mu m}^2$, acting as heat sinks. At the base temperature of 20 mK, the wire length of 50\,$\mu$m is between the inelastic electron scattering length of about 20\,$\mu$m and the electron-phonon interaction length of approximately 580\,$\mu$m \cite{henny1998}. Therefore, the wire is in the hot-electron regime, where electrons get heated inside the wire. The current noise is measured with a signal and spectrum analyser (SSA) at the same center frequency $f_0=2.9218$\,GHz and with the same bandwidth $\mathrm{BW} = 20$\,MHz as used for all the data in the main text. 

Fig.~\ref{figS2}(b) shows the shot noise dependence on current. The current noise spectral density $S_I$ is derived from the integrated noise power $\langle \delta P \rangle$ by subtracting the background noise at zero bias $\langle \delta P_0 \rangle$, dividing by the BW and taking into account the voltage division between the sample with resistance $R$ and the detection line with impedance $Z_0$ (see eq.~\ref{eq_power} and eq.~\ref{eq_dP_no}):
\begin{equation}
S_I = \frac{\langle \delta P \rangle - \langle \delta P_0 \rangle}{\mathrm{BW}} \cdot \frac{1}{g} \cdot \frac{\left(Z_0 + R \right)^2}{Z_0 R}.
\end{equation}
The setup gain $g$ contains the amplifier gain of the cryogenic and the room-temperature amplifiers and the setup attenuation. It is determined in the following way: $g$ is the slope of the detected, amplified noise in the linear regime (red dotted line) divided by the well-established Fano factor of a wire in the hot-electron regime of $F=\sqrt{3} /4$ \cite{steinbach1996}. The Fano factor decrease at high bias voltage is due to electron cooling via phonons.

\begin{figure}[H]
\centering
\includegraphics{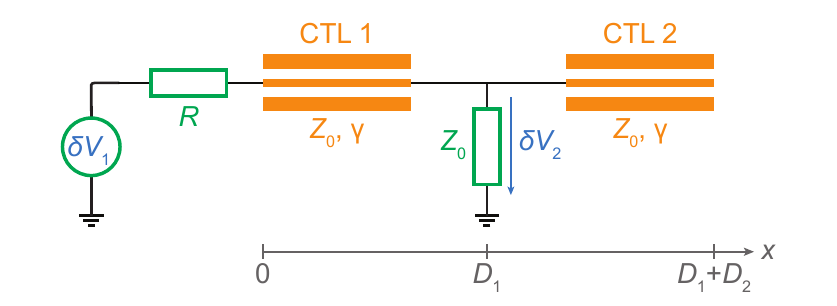}
\captionsetup{width=0.9\textwidth}
\caption{\label{figS3} Sketch of the stub tuner built by two CTLs of lengths $D_1$ and $D_2$. The noise source ($\delta V_1$) is the sample with differential resistance $R$ on the left side. The noise signal arriving at the detection line with characteristic impedance $Z_0$ is $\delta V_2$. }
\end{figure}

\textbf{Transmission function of a stub tuner.} 
Coplanar transmission lines (CTLs) are characterized by their characteristic impedance $Z_0$ and their complex propagation constant $\gamma=\alpha+\mathrm{i} \cdot \beta$. The real part of the propagation constant is the damping per length and the imaginary part the frequency-dependent wavenumber $\beta=2 \pi f \sqrt{\epsilon_\mathrm{eff}}/c$ with the effective dielectric constant $\epsilon_\mathrm{eff}$. A stub tuner is made out of two CTLs connected in parallel, as sketched in Fig.~\ref{figS3}. One of the CTLs with the length $D_1$ is terminated by the device with differential resistance $R$, whereas the other CTL with the length $D_2$ is open-ended. 

For the voltages in both CTL arms (CTL 1 and CTL 2), a wave function Ansatz is taken:
\begin{equation}
\begin{aligned}
V_1(x) &= V_1^{+} e^{-\gamma x} + V_1^{-} e^{\gamma x} \\
V_2(x) &= V_2^{+} e^{-\gamma x} + V_2^{-} e^{\gamma x}, 
\end{aligned}
\end{equation}
where $V_i^{+}$ and $V_i^{-}$ ($i=1,2$) are the coefficients for right-moving waves and left-moving waves, respectively. With the definition of the characteristic impedance $Z_0=V_i^{+}/I_i^{+}=-V_i^{-}/I_i^{-}$, one can write the current in the CTLs as
\begin{equation}
\begin{aligned}
I_1(x) &= \frac{1}{Z_0} \left( V_1^{+} e^{-\gamma x} - V_1^{-} e^{\gamma x} \right) \\
I_2(x) &= \frac{1}{Z_0} \left( V_2^{+} e^{-\gamma x} - V_2^{-} e^{\gamma x} \right). 
\end{aligned}
\end{equation}

To determine the four coefficients $V_i^{+}$ and $V_i^{-}$, four boundary conditions are needed. First, we require that the current at the open end vanishes, 
\begin{equation}
I_2(D_1+D_2)=0.
\end{equation}
The voltage on the other end is set by Ohm's law to
\begin{equation}
V_1(0)=\delta V_1-R I_1(0).
\end{equation}
Furthermore, the voltage has to be continuous at the connection of the two arms, 
\begin{equation}
V_1(D_1)=V_2(D_1).
\end{equation}
At last, Kirchhoff's law applies at the junction between the two arms:
\begin{equation}
I_1(D_1)=I_2(D_1)+\frac{V_1(D_1)}{Z_0}.
\end{equation}

What remains is a rather lengthy calculation to find the coefficients, which are used to evaluate the transmission function defined as
\begin{equation}
t_V = \frac{V_1(D_1)}{\delta V_1} = \frac{V_1^{+} e^{-\gamma D_1} + V_1^{-} e^{\gamma D_1}}{\delta V_1}.
\end{equation}
Eventually, the voltage-transmission function of a stub tuner is found to be
\begin{equation}
\label{eq_tv}
t_V = \frac{Z_0}{R+Z_0} \cdot \frac{2 e^{\gamma D_1} \coth \left(\gamma D_2 \right)}{\Gamma_L + e^{2 \gamma D_1} \left[ 1 + 2 \coth \left( \gamma D_2 \right) \right]},
\end{equation}
where the reflection coefficient $\Gamma_L = \left( R-Z_0 \right) /  \left( R+Z_0 \right)$. Around the resonance frequency $f_0=c/ \left( 2 \sqrt{\epsilon_\mathrm{eff}} \left( D_1+D_2 \right) \right)$, the stub tuner has a window of high transmission. In the lossless case ($\alpha=0$) and if $R \gg Z_0$, the condition for matching is $(D_1-D_2)/(D_1+D_2)=2/\pi \cdot \sqrt{Z_0/R}$ \cite{pozar2005}. If in addition only frequencies around $f_0$ are considered ($\Delta f \ll f$ with $\Delta f=f-f_0$), the magnitude square of the transmission function can be approximated to
\begin{equation}
\label{eq_tv_app}
\lvert t_V\rvert ^2 \approx \frac{1}{\pi ^2} \cdot \left( \frac{Z_0}{R} \right) ^3 \cdot \frac{1}{\left( \frac{2}{\pi} \frac{Z_0}{R} \right) ^2 + \left( \frac{\Delta f}{f_0} \right)^2}.
\end{equation}
The stub tuner bandwidth defined as full width at half maximum (FWHM) is in this ideal case 
\begin{equation}
\label{eq_BW_stub}
\mathrm{BW_{stub}}=\frac{4}{\pi} \cdot f_0 \cdot \frac{Z_0}{R}.
\end{equation}

\begin{figure}[H]
\centering
\includegraphics{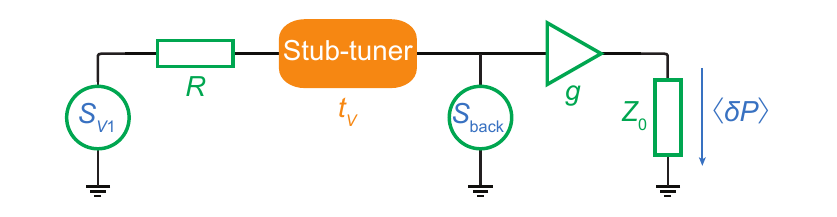}
\captionsetup{width=0.9\textwidth}
\caption{\label{figS4} Schematic of the measurement setup. The noise generated by the sample with differential resistance $R$ is drawn as a voltage source $S_{V_1}$ in series. The noise signal is transmitted through the stub tuner and amplified by a factor $g$. The instrument measures the integrated noise power $\langle \delta P \rangle$ over $Z_0$. The background noise ($S_{bg}$) is assumed to be added between the stub tuner and the amplifier. }
\end{figure}

\vspace{1em}
\textbf{Signal-to-noise ratio.} 
Given the transmission function of the stub tuner, one can quantify how much noise power arrives at the detection line. Fig.~\ref{figS4} shows a schematic of the situation. We assume the sample to emit a frequency-independent current noise spectral density $S_I$. Thus the voltage noise in front of the stub tuner is $S_{V_1}=S_I R^2$. The sum of the voltage noise integrated over the measurement bandwidth $\mathrm{BW}$ after passing the stub tuner and the amplifier with power gain $g$ is $\langle \delta V_3^2 \rangle = S_{V_1} g \int_\mathrm{BW_{stub}} \lvert t_V\rvert ^2 \, df$. Finally, the noise power measured over a resistance of $Z_0$ is $\langle \delta P \rangle = \langle \delta V_3^2 \rangle / Z_0$. Putting everything together results in the measured noise power
\begin{equation}
\label{eq_power}
\langle \delta P \rangle = S_I \frac{R^2}{Z_0} g \int_\mathrm{BW} \lvert t_V\rvert ^2 \, df .
\end{equation}
This is eq.~4 in the main text without taking the background noise into account yet. 

While the general integration of the stub tuner transmission function (eq.~\ref{eq_tv}) has to be done numerically, an analytical expression exists for the integral of the lossless transmission function at matching (eq.~\ref{eq_tv_app}):
\begin{equation}
\label{eq_tv_int}
\int_{-\infty}^{\infty} \lvert t_V\rvert ^2 \, df = \frac{1}{2} f_0 \left( \frac{Z_0}{R} \right)^2.
\end{equation}
It defines an upper bound for the detectable noise power with a stub tuner, which is
\begin{equation}
\label{eq_dP_max}
\langle \delta P \rangle_\mathrm{stub}^\mathrm{max} = \frac{1}{2} S_I Z_0 g f_0.
\end{equation}

This transmitted noise power has to be compared with the signal measured in the absence of impedance matching. The noise spectral density is small in this case, but one can integrate over a large bandwidth. On the other hand, the bandwidth is always restricted by other components. In our setup, the circulator has the smallest bandwidth of $\mathrm{BW_0}=500$ MHz. Eq.~\ref{eq_power} can be generally used to get the signal power for this kind of setup by taking the appropriate transmission function $t_V$. If there is no impedance-matching circuit at all, the stub tuner is replaced by an element with a constant transmission function $t_V=Z_0/(Z_0+R)$ obtained via eq.~\ref{eq_tv} by setting $D_1=D_2=0$. It leads to the same result as if voltage division of the emitted noise voltage is considered in the circuit drawn in Fig.~\ref{figS4}. By means of eq.~\ref{eq_power}, the resulting noise power measured without impedance matching is then
\begin{equation}
\label{eq_dP_no}
\langle \delta P \rangle_0 = S_I Z_0 g \mathrm{BW_0}.
\end{equation}
Therefore, the maximum enhancement in detectable noise power obtained with a lossless stub tuner at full matching at a resonance frequency $f_0=3$ GHz is $\langle \delta P \rangle_\mathrm{stub}^\mathrm{max} / \langle \delta P \rangle_0 = 1/2 \cdot f_0/\mathrm{BW_0} \approx 3$. 

So far the discussion was about the noise \textit{signal} only. But the main advantage of a matching circuit is revealed by considering the signal-to-noise ratio (SNR). Noise is in this context the background noise, as for instance the amplifier noise. Its power spectral density $S_{bg}$ is assumed to be frequency independent. If one assumes the background noise source to be introduced between the impedance-matching circuit and the amplifier (see Fig.~\ref{figS4}), the background noise power picked up over the measurement bandwidth (BW) is $\langle \delta P \rangle_{bg}=S_\mathrm{bg} \mathrm{BW} g$. This leads to the general expression for the SNR
\begin{equation}
\label{eq_SNR}
\mathrm{SNR} = \frac{\langle \delta P \rangle}{\langle \delta P \rangle_{bg}} = \frac{S_I R^2}{S_{bg} Z_0} \cdot \frac{\int_\mathrm{BW} \lvert t_V\rvert ^2 \, df}{\mathrm{BW}}, 
\end{equation}
where eq.~\ref{eq_power} is used to get the last expression. Without impedance matching, one can use eq.~\ref{eq_dP_no} and gets
\begin{equation}
\label{eq_SNR_0}
\mathrm{SNR_0} = \frac{\langle \delta P_0 \rangle}{\langle \delta P \rangle_{bg}} = \frac{S_I Z_0}{S_{bg}}.
\end{equation}

Eventually, we want to compare the SNRs with and without impedance matching. To do so, we introduce the figure of merit for impedance matching as
\begin{equation}
\label{eq_g}
g_\mathrm{SNR} = \frac{\mathrm{SNR}}{\mathrm{SNR_0}} = \left( \frac{R}{Z_0} \right) ^2 \cdot \frac{\int_\mathrm{BW} \lvert t_V\rvert ^2 \, df}{\mathrm{BW}}.
\end{equation}
This means that $g_\mathrm{SNR}$ depends on the transmission function $t_V$ of the resonant circuit and the chosen integration bandwidth BW, which is optimally the FWHM given in eq.~\ref{eq_BW_stub}. An upper bound for $g_\mathrm{SNR}$ considering a lossless stub tuner at full matching can then be given with the help of eq.~\ref{eq_tv_int}:
\begin{equation}
\label{eq_SNR_stub}
g_\mathrm{SNR}^\mathrm{max} = \frac{\pi}{8} \frac{R}{Z_0},
\end{equation}
which amounts to a factor as high as $g_\mathrm{SNR} \approx 800$ for $R=100$\,k$\Omega$. For realistic matching circuits, the integration of the general transmission function (eq.~\ref{eq_tv}) can be done numerically. Using the stub-parameters from the main text derived by reflectometry, the resulting improvement in the SNR is still $g_\mathrm{SNR} \approx 200$ at this resistance if the bandwidth is the FWHM, although the circuit is not fully matched to $R=100$\,k$\Omega$.

\begin{figure}[H]
\centering
\includegraphics{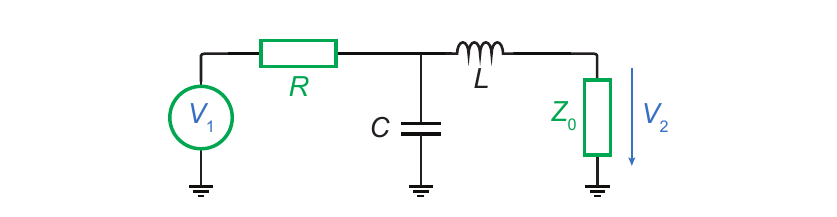}
\captionsetup{width=0.9\textwidth}
\caption{\label{figS5} Circuit diagram for an $LC$ impedance-matching circuit. The voltage $V_1$ present at the sample with resistance $R$ is transformed to the voltage $V_2$ in the detection line with characteristic impedance $Z_0$.}
\end{figure}

\vspace{1em}
\textbf{Comparison with $LC$-circuit}
In the last part, we compare this result with an $LC$ matching circuit. To this end, an expression for the voltage-transmission function $t_V$ appearing in eq.~\ref{eq_g} has to be derived. Using the circuit drawn in Fig.~\ref{figS5}, the transmission function is determined to be
\begin{equation}
\label{eq_tv_lc}
t_V = \frac{V_2}{V_1} = \frac{Z_0}{R} \cdot \frac{1}{1+\frac{Z_0}{R}+\mathrm{i} \cdot 2 \sqrt{\frac{Z_0}{R}} \left( 1+\frac{\Delta f}{f_0} \right) - \left( 1+\frac{\Delta f}{f_0} \right)^2},
\end{equation}
at matching, when $L/C=R Z_0$. Here $\Delta f=f-f_0$. In the limit where $R \gg Z_0$ and $\Delta f \ll f_0$, $t_V$ can be approximated and its magnitude squared reads
\begin{equation}
\label{eq_tv_lc_app}
\lvert t_V\rvert ^2 \approx \frac{1}{4} \cdot \left( \frac{Z_0}{R} \right)^2 \cdot \frac{1}{\frac{Z_0}{R} + \left( \frac{\Delta f}{f_0} \right)^2}. 
\end{equation}

The FWHM of this resonance function is $\mathrm{BW}_{LC}=2f_0\sqrt{Z_0/R}$. It scales with $\sqrt{Z_0/R}$ whereas the bandwidth of the stub tuner scales with $Z_0/R$ (eq.~\ref{eq_BW_stub}), leading at high resistances $R$ to a much larger BW for an $LC$ circuit compared with a stub tuner. Furthermore, the integral of the transmission function in eq.~\ref{eq_tv_lc_app} can be evaluated analytically:
\begin{equation}
\label{eq_int_tv_lc}
\int_{-\infty}^{\infty} \lvert t_V\rvert ^2 \, df = \frac{\pi}{4} f_0 \left( \frac{Z_0}{R} \right)^{3/2}. 
\end{equation}
What remains is to plug $\mathrm{BW}_{LC}$ and the integrated transmission function into eq.~\ref{eq_g} in order to get the maximum figure of merit for a lossless, fully matched $LC$ network:
\begin{equation}
\label{eq_g_lc}
g_\mathrm{SNR}^\mathrm{max} = \frac{\pi}{8} \frac{R}{Z_0}. 
\end{equation}
It is exactly the same result as for the stub tuner (eq.~\ref{eq_SNR_stub}). This is not surprising since with any matching circuit, the transmission maximum is fixed to $t_V=1/4 \cdot Z_0/R$ and only the BW can be modified. Still, the large bandwidth of an $LC$ matching circuit can be beneficial for some measurements, for example a fast, time-resolved read-out. But the gain in the SNR remains the same, since the increase in integration bandwidth also leads to an enhanced background noise. 

In conclusion, an impedance-matching circuit leads to a significant increase in the figure of merit. The discussion here concentrates on the improvement of the SNR with impedance matching. But we want to note that it is equally important to minimize the background noise $S_\mathrm{bg}$ in the first place. By looking at eq.~\ref{eq_SNR}, it gets evident that apart from impedance matching to achieve a high signal transmission and properly choosing the right BW, the only way to increase the SNR is to reduce $S_\mathrm{bg}$, for instance with Josephson parametric amplifiers. On the other hand, the SNR is not the only limiting characteristic of an experiment. What counts experimentally is to reach a certain SNR which is high enough for the intended accuracy. Any further increase of the SNR does not reveal new features. But the larger the figure of merit is, the smaller is the measurement time needed to reach the same SNR.